\documentclass[journal]{IEEEtran}
\IEEEoverridecommandlockouts
\usepackage{cite}
\usepackage{amsmath,amssymb,amsfonts}
\usepackage{graphicx}
\usepackage{textcomp}
\usepackage{xcolor}
\usepackage[utf8]{inputenc}
\usepackage{pgfplots} 
\pgfplotsset{compat=newest}
\pgfplotsset{plot coordinates/math parser=false}
\usetikzlibrary{plotmarks}
\newlength\figureheight
\newlength\figurewidth
\usepackage{helvet}
\usepackage[eulergreek]{sansmath}
\usepackage{algorithm,algpseudocode,float}
\usepackage{stackengine}
\usepackage{xurl}
\usepackage{caption}
\usepackage{comment}
\usepackage{amsthm}
\usepackage{booktabs}
\usepackage{comment}

\newcommand\simplex{{\cal S}}
\newcommand\model{{\cal M}}

\def\BibTeX{{\rm B\kern-.05em{\sc i\kern-.025em b}\kern-.08em
    T\kern-.1667em\lower.7ex\hbox{E}\kern-.125emX}}

\begin{document}

\title{Maximum entropy and quantized metric models for absolute category ratings
\thanks{Funded by the Deutsche Forschungsgemeinschaft (DFG, German Research Foundation) -- Project-ID 251654672 -- TRR 161}
\thanks{We gratefully acknowledge Polish high-performance computing infrastructure PLGrid (HPC Center: ACK Cyfronet AGH) for providing computer facilities and support within computational grant no.\ PLG/2023/016748. We would like to thank Bogdan \'Cmiel for sharing his expertise.}
}

\author{\IEEEauthorblockN{
Dietmar Saupe\IEEEauthorrefmark{1},
Krzysztof Rusek\IEEEauthorrefmark{2},
David Hägele\IEEEauthorrefmark{3},
Daniel Weiskopf\IEEEauthorrefmark{3},
Lucjan Janowski\IEEEauthorrefmark{2}}\\
\IEEEauthorblockA{\IEEEauthorrefmark{1}\textit{Department of Computer and Information Science, University of Konstanz, Konstanz, Germany } }
\IEEEauthorblockA{\IEEEauthorrefmark{2}\textit{AGH University of Science and Technology, Krakow, Poland}} 
\IEEEauthorblockA{\IEEEauthorrefmark{3}\textit{VISUS, University of Stuttgart, Stuttgart, Germany}} 
dietmar.saupe@uni-konstanz.de, 
\{krusek,lucjan.janowski\}@agh.edu.pl,\\
\{david.haegele,daniel.weiskopf\}@visus.uni-stuttgart.de}

\maketitle

\begin{abstract}
The datasets of most image quality assessment studies contain ratings on a categorical scale with five levels, from bad (1) to excellent (5).
For each stimulus, the number of ratings from 1 to 5 is summarized and given in the form of the mean opinion score. 
In this study, we investigate families of multinomial probability distributions parameterized by mean and variance that are used to fit the empirical rating distributions. 
To this end, we consider quantized metric models based on continuous distributions that model perceived stimulus quality on a latent scale. 
The probabilities for the rating categories are determined by quantizing the corresponding random variables using threshold values. 
Furthermore, we introduce a novel discrete maximum entropy distribution for a given mean and variance.
We compare the performance of these models and the state of the art given by the generalized score distribution for two large data sets, KonIQ-10k and VQEG HDTV.
Given an input distribution of ratings, our fitted two-parameter models predict unseen ratings better than the empirical distribution. 
In contrast to empirical ACR distributions and their discrete models, our continuous models can provide fine-grained estimates of quantiles of quality of experience that are relevant to service providers to satisfy a target fraction of the user population.
\end{abstract}


\section{Introduction}
\label{sec_introduction}
\IEEEPARstart{I}{n} 
general, image processing applications with the goal of outputting images for human consumption can benefit from models of perceived image quality. Often, such models are created by machine learning, which typically requires training sets of images with annotations of perceived quality. To this end, human subjects are presented with images and asked to rate their visual quality, usually on an absolute category rating (ACR) scale, i.e., they are asked to select a quality from the scale $\Omega = \{1,2,3,4,5\}$. However, subjects might disagree, and so many ratings are collected for each image. The collected distribution of ratings for an image is summarized by the mean opinion score (MOS), i.e., the average rating \cite{ITU-R_BT.500}.

The ratings for a stimulus are responses from a sample of the population of potential media consumers. Our research question is how to estimate a model for the distribution in the entire population. 
The most obvious model candidate is the empirical one, given by the frequency of ratings in the sample itself.
Our main goal is to obtain models that provide a better fit to the population-wide distribution of ratings. 

As our first contribution, we propose using the class of quantized metric models as a general methodology. It works in two stages. First, the common psychological model of a latent quality scale is adopted. This assumes that subjects perceive the quality $Y$ of a stimulus on an interval scale following a certain probability distribution function (PDF). Usually, this interval is assumed to be the entire real line 
and the PDF of $Y$ is the normal distribution. For generalized and improved models, we also consider bounded interval scales $[0,1]$ and other types of parameterized PDFs than the normal distribution.

In the second stage, subjects produce responses by mapping their perceived quality values to the discrete, finite ACR scale. This is most easily modeled by a quantization function. The interval scale is partitioned into five intervals by certain threshold values $\tau_k$. The discrete distribution model then prescribes the probability $\Pr(X=k)$ as the probability that $Y$ lies in the $k$-th interval. The random variable $X$ is a quantized version of~$Y$. The threshold values can either be fixed or, to achieve better performance, selected adaptively using an adjustment procedure.

As our second contribution, we introduce parameterized maximum entropy distributions that directly model discrete ACR distributions. This novel model focuses on exactly satisfying all constraints defined by their parameters (mean and variance) while assuming maximal uncertainty about all other details in the distribution. 

For both of the above approaches---quantized metric models and maximum entropy distributions---the number and types of model parameters must be carefully selected. Their number determines the dimensionality of the search space for a model of a given empirical distribution of ratings. There are four degrees of freedom in the space of probability distributions for the ACR scale. Each parameter reduces this dimensionality by one. Previous work of Hoßfeld et al.~\cite{hossfeld2016qoe} has shown that the mean as a single parameter provides insufficient information about the distribution for the purpose of, e.g., understanding the uncertainty of user ratings or the ratio of satisfied users. However, for one particular case, taking the mean together with the standard deviation of opinion scores, entire distributions could be accurately estimated \cite{hossfeld2020from}. Motivated by these results, our models of ACR distributions also have two parameters defining their mean and variance. Our results confirm that already these two parameters are sufficient to provide statistically valid models. 
    
The discussion of model performance focuses on two main features: fit and prediction. Model fits are compared using standard statistical likelihood ratios in the form of the G-test and p-values. To quantify the predictive power of our models, we bootstrap random samples of up to 40 ratings. We show that our models derived from these samples provide a better fit to the remaining test data than the empirical distributions from the sample itself. Furthermore, we estimate the number of additional ACR ratings that would be required to predict directly from the data in order to reach the performance of our model predictions.

The state of the art of modeling and representing distributions of ACR quality assessments can be summarized as follows. Seufert~\cite{seufert2021statistical} discussed the fundamental advantages of distributions over MOS-based metric models and recommended the empirical multinomial distribution model. Hoßfeld et al.~\cite{hossfeld2016qoe} also argued for going beyond MOS when performing subjective quality tests, for example, by a binomial model of ACR distributions. Liddell and Kruschke \cite{liddell2017analyzing} showed how quantized metric models can overcome the limitations of metric models applied to ordinal data. Only a few instances of quantized metric models were reported for quality assessment \cite{van1995quality, tasaka2017bayesian, pezzulli2020estimation, saupe2024national}. However, none of these works compared the performance of quantized metric models (and others) as in our paper.

Our contribution was inspired by the generalized score distribution (GSD) \cite{GSDTOM}, also a distribution with parameters for mean and variance. It has been shown to be a statistically valid model for ACR distributions derived from properly conducted subjective quality assessment studies. It was compared with two versions of the quantized normal metric model and performed best of the three models in terms of fit. Our analysis includes the GSD as a candidate model and compares its performance. We are unaware of any other work that aims to uncover the hidden structures of ACR rating distributions.

\section{Models}

In this section, we propose different models with a common parametrization as follows. Image quality assessment using the 5-level ACR scheme collects ratings ranging over the scale $\Omega = \{1,\ldots,5\}$. Dividing the counts for each level gives an (empirical) probability mass function (PMF). It is represented by a vector $p \in \simplex$, where $\simplex = \{p\in[0,1]^5~|~\sum_1^5 p_k = 1\}$, the standard simplex in $\mathbb{R}^5$. For this PMF $p$, we denote its mean (MOS) by $\psi = \Psi(p)$. 

For a PMF with mean $\psi$, the variance $v$ ranges over an interval $[v_{\min},v_{\max}]$ \cite{hossfeld2011sos}:
\begin{align}
   \nonumber v_{\min} &= (\lceil \psi \rceil-\psi)(\psi-\lfloor \psi \rfloor) \in [0,0.25],\\ 
    \nonumber v_{\max} &= (\psi-1)(5-\psi) \in [0,4].
\end{align}
The complementary normalized variance for a PMF $p$ is 
$$
    \rho = \varrho(p) = \frac{v_{\max}-v}{v_{\max}-v_{\min}} \in [0,1]
$$
and ranges from the maximal variance ($\rho=0$) to minimal variance ($\rho=1$). For $\mu = 1$ and $\mu = 5$, $v_{\min} = v_{\max} = 0$, and $\rho$ is undefined and irrelevant.
Our model distributions are parameterized by the mean $\psi$ and variance $v$ or the complementary normalized variance $\rho$.

\subsubsection{Quantized metric models}
\label{sec_QMM}
Metric models assume a real random variable $Y$ with a PDF $f$ for the perceived quality on the latent scale. For example, in ITU P.913, normal distributions are assumed underlying the ratings \cite{ITU-T_P.913}. However, such models must be quantized to produce discrete probability distributions for the five categories in the ACR scheme. Thus, for the discrete random variable $X$ that models ACR with 5 levels, we assume thresholds \mbox{$\tau_0 = -\infty < \tau_1 < \cdots  < \tau_4 < \tau_5 = \infty$} and set
$
\Pr(X = k) = F(\tau_k)-F(\tau_{k-1}),
$
where $F$ denotes the cumulative probability function (CDF) corresponding to the PDF $f$. 

As a probability distribution, we selected the normal distribution, defined on the real line, as it was recommended in ITU P.913, albeit without quantization. The logistic distribution is often used for modeling categorical dependent variables in latent variable models. It is very similar to the normal distribution but has heavier tails that can increase the robustness of analyses. For the normal or logistic distributions, it cannot be ruled out that some of the reconstructed stimuli qualities are outside the range $[1,5]$. This effect can be avoided by considering distributions defined on the unit interval $[0,1]$. As an example, we considered the logit-logistic distribution, which describes a random variable whose logit transform has a logistic distribution. Additionally, we included the beta distribution because it is the most commonly employed two-parameter distribution for modeling random variables on the unit interval \cite{smithson2017cdf} and has already been used for modeling ACR distributions \cite{hossfeld2020from}. 

Following Liddell and Kruschke~\cite{liddell2017analyzing} and Nawa{\l}a et al.~\cite{GSDTOM}, we have chosen the threshold values $\tau_1$ to  $\tau_4$ at 1.5, 2.5, 3.5, and 4.5 for the normal distribution. In this way, the expectation of $X$ is comparable to the MOS on the ACR scale. This is not possible for the other distributions defined on the unit interval. For those, we chose thresholds  0.2, 0.4, 0.6, and 0.8 such that the five successive intervals defining the probability mass for the five ACR categories are of equal length, namely 1/5.

\subsubsection{The maximum entropy model}
The principle of maximum entropy states that the probability distribution that best represents the current state of knowledge about a system is the one with largest entropy \cite{guiasu1985principle}.  Following the principle of maximum entropy, we define the maximum entropy PMF as
$$
    q_{\text{Hmax}}(\psi,\rho) = \arg \max_{q \in \simplex}~\{H(q)~|~\Psi(q)=\psi, \varrho(q) = \rho\},
$$
where $H(q)$ denotes the entropy of $q$. 
Informally, the maximum entropy distribution is the most uniform one, subject to the constraints of mean and variance. Note that the maximum entropy distribution is unique. This follows from the strict concavity of the entropy function and the convexity of the set of probability distributions with fixed mean and variance.

The normal distribution is the corresponding continuous univariate probability distribution that has maximum entropy for fixed mean and variance. In this sense, the discrete distribution with maximum entropy has more in common with the normal distribution than the quantized normal distribution.

\begin{table}[t]
\centering
\caption{Goodness of fit for dataset KonIQ-10k.}
\resizebox{1.0\columnwidth}{!}{%
\small
\begin{tabular}{cr cccc}  
\toprule
Rank  &    &  AIC $\downarrow$  & G-test $\downarrow$ & Ratio \\ 
(G-test) &  Model  & $(\times 10^6)$  & Mean & $p<0.05$ \\ 
\midrule
1 &  logit-logistic  & 1.876$\pm$0.005  & 1.692$\pm$0.035 &  0.0316\\ 
2 &      maxentropy  & 1.877$\pm$0.005  & 1.790$\pm$0.047 &  0.0594\\ 
3 &          normal  & 1.878$\pm$0.005  & 1.901$\pm$0.059 &  0.0725\\ 
4 &            beta  & 1.878$\pm$0.005  & 1.908$\pm$0.051 &  0.0755\\ 
5 &        logistic  & 1.879$\pm$0.005  & 1.968$\pm$0.040 &  0.0508\\ 
6 &             GSD  & 1.903$\pm$0.005  & 4.428$\pm$0.077 &  0.2773\\ 
\bottomrule
\end{tabular}
}
\label{table_KonIQ}
\end{table}

\subsubsection{Generalized score distribution model (GSD)}
The GSD is a class of modified binomial distributions covering the entire set of possible means and variances. For underdispersed cases, it is a mixture of binomial and Bernoulli distributions; for overdispersed cases, it is a reparametrized beta-binomial distribution. Its parameters are $\psi$ and $\rho$. For the details, see Nawa{\l}a et al.~\cite{GSDTOM} and Ćmiel et al.~\cite{GSDSP}.

\subsubsection{Empirical frequency distribution}
The empirical frequency distribution is obtained from a sample of ACR ratings by normalizing the frequencies of the rating categories in $\Omega = \{1,2,3,4,5\}$. The Fundamental Theorem of Statistics states that the corresponding cumulative empirical distribution function converges uniformly to the true distribution function almost surely \cite{vaart1998asymptotic}.  We compare the empirical model with ours in terms of the accuracy of the predicted true distributions. While our models have only two degrees of freedom (mean and variance), the specification of an empirical model requires four parameters.

\section{Methods}
\label{sec_methods}

We selected two large-scale datasets with diversity in stimuli type (images and videos) and acquisition mode (crowdsourcing and lab). KonIQ-10K contains 10,073 ACR distributions with 1.07 million ratings for image quality (107 per image on average), collected by crowdsourcing from 1261 subjects  \cite{hosu2020koniq}. VQEG-HDTV is a video quality dataset acquired in a lab study with 24 ratings for each of the 864 videos \cite{HDTV_Phase_I_test}.

\subsection{Maximum likelihood estimation and goodness of fit}
Each of the models from above defines a set $\model$  of PMFs, parametrized by two parameters, for example, by the mean and variance in the quantized metric model from the normal distribution or $\psi$ and $\rho$ for the maximum entropy and GSD models. For a given model and empirical PMF $p \in \simplex$, we must find the best-matching model instance $q \in \model$.
For this purpose, we applied maximum likelihood estimation (MLE) by minimizing the cross-entropy  $H(p,q) = - \sum_{k=1}^K p_k {\log q_k}$ over all $q \in \model$. The optimizations were done using an interior point method as implemented in the Matlab function \textsc{Fmincon} \cite{fmincon}.

To assess the goodness of fit of the models for our data sets, we calculated the Akaike information criterion (AIC) and the likelihood ratio given by the G-test statistic  averaged over all distributions. 95\% confidence intervals were estimated from 1000 bootstrap samples.

The G-test is a likelihood-ratio test that allows us to check whether the number of observed ratings in each category fits the theoretical expectation given by each of our models \cite{Agresti}. For selecting a model among a list of candidates, Akaike's information criterion (AIC) is among the most popular and versatile strategies \cite{claeskens2008model}. AIC rewards goodness of fit (as assessed by the likelihood function) but also includes a penalty that is proportional to the number of estimated parameters.

\subsection{Prediction accuracy for unseen data}
Models with a good fit do not necessarily produce reliable predictions of unseen data. Therefore, we also assessed the prediction accuracy in the second part of our data analysis.

While the goodness of fit is evaluated using the training data, the evaluation of the prediction performance requires new data (or cross-validation). The KonIQ-10k dataset is sufficiently large, with an average of 107 ratings per stimulus that can be divided into a training and a test set. We estimated the predictive performance of the logit-logistic and GSD model and compared it with that of the empirical model.

For each trial, a random image was selected from KonIQ-10k and a sample of $n$ ratings was taken for training. The remaining ratings served as test data. For each sample, we computed the models and compared them to the distribution of the test data. The following metrics were applied: Maximum ($L^{\infty}$), Euclidean, Bhattacharyya, Kolmogorov-Smirnov, and Wasserstein. For the sample sizes $n = 10, 11, \ldots, 40$, the mean values from 10,000 trials were calculated.

\begin{figure*}[t]
\centering
\includegraphics[width=0.32\linewidth]{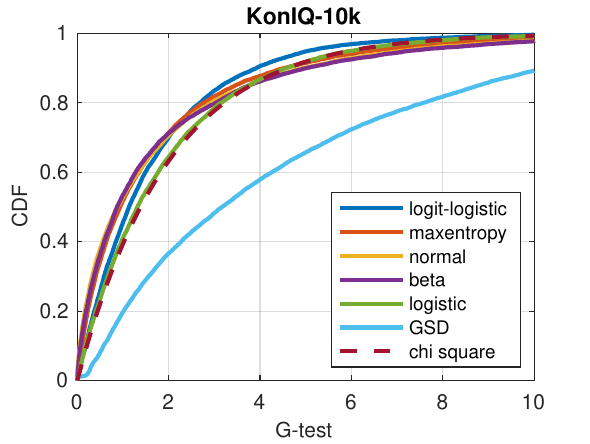}
\includegraphics[width=0.32\linewidth]{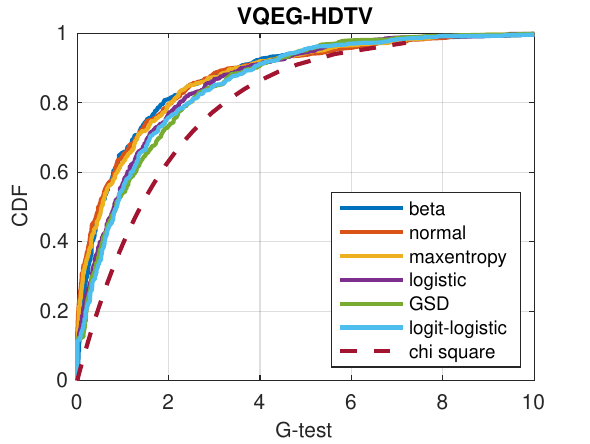}
\includegraphics[width=0.32\linewidth]{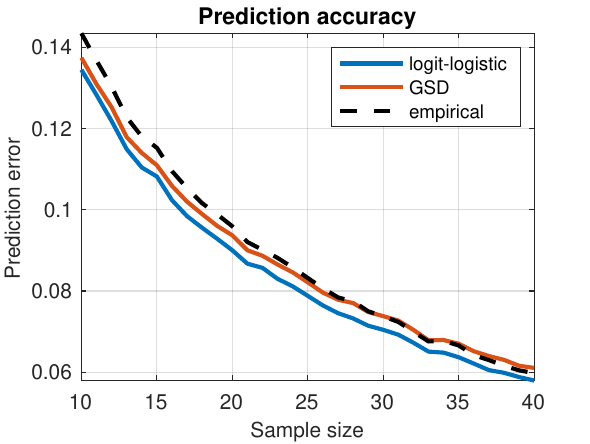}
\caption{The goodness of the model fit is shown by the G-test CDFs  (left and center). Curves above the dashed line for the asymptotic chi-squared distribution indicate a statistically valid fit. The right figure shows that the logit-logistic and the GSD models predict unseen data better than the empirical model from samples of 10 to 40 ACR ratings of KonIQ-10k. The prediction errors are given as $L^{\infty}$-distances of the corresponding distributions (see Table \ref{table_pred_logistic} for more details).}
\label{fig_cdfs}
\end{figure*}

\begin{table}[t]
\centering
\caption{Goodness of fit for dataset VQEG HDTV.}
\resizebox{1.0\columnwidth}{!}{%
\small
\begin{tabular}{cr cccc}  
\toprule
Rank  &    &  AIC $\downarrow$  & G-test $\downarrow$ & Ratio \\ 
(G-test) &  Model  & $(\times 10^4)$  & Mean & $p<0.05$ \\ 
\midrule
1 &            beta  & 4.401$\pm$0.071  & 1.237$\pm$0.115 &  0.0324\\ 
2 &          normal  & 4.402$\pm$0.069  & 1.244$\pm$0.132 &  0.0428\\ 
3 &      maxentropy  & 4.403$\pm$0.069  & 1.261$\pm$0.126 &  0.0359\\ 
4 &        logistic  & 4.415$\pm$0.071  & 1.400$\pm$0.115 &  0.0278\\ 
5 &             GSD  & 4.421$\pm$0.071  & 1.470$\pm$0.110 &  0.0208\\ 
6 &  logit-logistic  & 4.422$\pm$0.071  & 1.481$\pm$0.124 &  0.0301\\ 
\bottomrule
\end{tabular}
}
\label{table_VQEG}
\end{table}

\section{Results and discussion}
\label{sec_results}
The results for the goodness of fit are presented in Tables~\ref{table_KonIQ} and \ref{table_VQEG} 
and show that all models perform similarly on both datasets, except for the GSD on KonIQ-10k. 

The last columns in the tables give the ratios of the distributions with a p-value of the G-test statistic less than 0.05. Such small ratios indicate ACR rating distributions for which the corresponding model should be rejected at the 95\% confidence level. The p-values were calculated using the asymptotic estimate given by the chi-squared distribution with two degrees of freedom \cite{mcdonald2009handbook}. Under the null hypothesis that the model is representative of the data, the expected ratio of p-values less than 0.05 equals 0.05. Thus, ratios of less than 0.05 do not justify rejection of the null hypothesis. Based on this criterion,  the logit-logistic model passes the test for KonIQ-10k, and for VQEG HDTV, all of the methods pass the test. 

This result confirms that for these two datasets, two parameters can provide statistically valid models for the distributions. The same conclusion is suggested by the principal component analysis (PCA) for the two datasets of empirical ACR distributions. We found that already 95.5\% and 83.3\% of the variances of the distributions in KonIQ-10k and VQEG HDTV, respectively, are explained by the first two components.

Figure \ref{fig_cdfs} (left and center) shows the CDFs of the G-test statistics together with the asymptotic chi-squared distribution (df $= 2$). Larger cumulative densities indicate better goodness of fit. For an acceptable fit, the CDF of the model should be above or close to the chi-squared CDF. The diagrams confirm the above findings obtained from the table values. 

\begin{table}[t]
\centering
\caption{Prediction accuracy for the logit-logistic model for dataset KonIQ-10k.}
\resizebox{1.0\columnwidth}{!}{%
\small
\begin{tabular}{cccccc}  
\toprule
Sample  & Model &  Model  & Effect & Gain  \\ 
size &  log.-logistic  & empirical  & size &  \\ 
ratings &  $L^{\infty}$-error $\downarrow$ & $L^{\infty}$-error $\downarrow$ & Cohen's $d \uparrow$ & samples $\uparrow$ \\ 
\midrule
        10  &        0.135 &    0.144 &   1.39 &   1.35 \\
        15  &        0.108 &    0.115 &   1.67 &   1.32 \\
        20  &        0.090 &    0.096 &   1.97 &   2.05 \\
        25  &        0.079 &    0.083 &   1.97 &   1.83 \\
        30  &        0.070 &    0.074 &   1.88 &   1.80 \\
        35  &        0.064 &    0.067 &   2.00 &   1.42 \\
        40  &        0.058 &    0.060 &   1.54  &  1.22\\ 
\bottomrule
\end{tabular}
}
\label{table_pred_logistic}
\end{table}

Table \ref{table_pred_logistic} compares the prediction accuracy for the logit-logistic and the empirical model. The errors are expressed as $L^{\infty}$-distances between the model distributions derived from the samples and the empirical distributions of ratings in the test sets. Figure \ref{fig_cdfs} (right) shows the mean values of the $L^{\infty}$-metric, including those of the GSD models. The prediction accuracy for the logit-logistic model is better than for the empirical model. The effect sizes are ``very large'' according to Cohen's~$d$. For the other metrics, the results are very similar.

The last column in the table shows the ``gain,'' defined as the additional number of ratings that the empirical model would need to achieve the predictive accuracy of the logit-logistic model. For example, for samples of 20 ratings, the gain is about two ratings. Thus, if an empirical model is desired with the same prediction accuracy as the logit-logistic quantized metric model derived from samples of 20 ratings, then samples of at least 22 ratings must be acquired for the empirical model. As shown in Figure~\ref{fig_cdfs} (right), these gains correspond to the amount of left shift applied to the graph for the empirical model to match the graph for the logit-logistic model.

The modeling of distributions of ratings from samples of small size like 10 to 40 opinions (Table \ref{table_pred_logistic}) can be interpreted as a (parametric) statistical smoothing method \cite{simonoff2012smoothing}. The weak assumption of smoothness of the perceived stimulus quality on the continuous latent scale is reasonable. However, with small sample sizes, noisy data is introduced that limits the fitting of the empirical distribution and its prediction power. Smoothing can help, for example, by allowing the assignment of non-zero probabilities to bins that do not occur in the sample.

\section{Conclusion}
We investigated how to best model the distributions of ACR image and video quality ratings.  Our models are successful with acceptable goodness of fit, can predict unseen data better than the empirical distributions, and are scalable to absolute category ratings with more than five levels.

Given a distribution of ratings, the question remains which one of the models to apply. We recommend choosing and testing different model candidates and selecting the one with the best fit or the greatest prediction power in a cross-validation. For this purpose, we provide code and a more detailed comparison of the individual models along with further tables and visualizations \cite{OurGitHub}. 
The models can be applied for the analysis of ACR data. For example, quantiles of opinion scores are  relevant for service providers but can only be given coarsely, as integers 1 to 5, when using the empirical distributions \cite{hossfeld2016qoe}. On the continuous latent scale of the models, the quantiles vary continuously, providing fine-grained results.

Another application is the parametric bootstrap. 
The nonparametric bootstrap may perform poorly for small sample sizes because empty cells and spurious fine structure may not be present in the population sampled, whereas in the parametric bootstrap, these irregularities are smoothed out.

\newpage
\bibliographystyle{unsrt}
\bibliography{mainrefs}

\end{document}